\documentclass[notitlepage,superscriptaddress,showpacs,nobalancelastpage,twocolumn,aps,prl]{revtex4-1}

\usepackage[english]{babel}
\RequirePackage[T1]{fontenc}
\RequirePackage{times} 

\usepackage{siunitx} 
\usepackage[italicdiff]{physics}
\usepackage{amsfonts}
\usepackage{subfigure}
\usepackage{amsmath}
\usepackage{amssymb}
\usepackage{graphicx,epstopdf}
\usepackage{xspace}
\usepackage{array}
\usepackage[hidelinks]{hyperref}
\usepackage{xcolor}
\usepackage{my_symbols}
\usepackage{CJK}
\usepackage{bm}
\usepackage{verbatim}
\usepackage{tikz}
\usetikzlibrary{calc,3d}
\usetikzlibrary{arrows}
\usetikzlibrary{snakes}
\usepackage[normalem]{ulem}

{\par}

\newcounter{mycomment}

\newcommand\rmv{\bgroup\markoverwith {\textcolor{red}{\rule[0.5ex]{2pt}{0.4pt}}}\ULon}

\begin{document}
\begin{CJK*}{UTF8}{gbsn} 
\title{Realization of Attractive Level Crossing via a Dissipative Mode}
\author{Weichao Yu (余伟超)}
\affiliation{Department of Physics and State Key Laboratory of Surface Physics, Fudan University, Shanghai 200433, China}
\affiliation{Institute for Materials Research, Tohoku University, Sendai 980-8577, Japan}
\author{Jiongjie Wang}
\affiliation{Department of Physics and State Key Laboratory of Surface Physics, Fudan University, Shanghai 200433, China}
\author{H. Y. Yuan}
\affiliation{Department of Physics, Southern University of Science and Technology, Shenzhen 518055, Guangdong, China}
\author{Jiang Xiao (萧江)}
\email[Corresponding author:~]{xiaojiang@fudan.edu.cn}
\affiliation{Department of Physics and State Key Laboratory of Surface Physics, Fudan University, Shanghai 200433, China}
\affiliation{Institute for Nanoelectronics Devices and Quantum Computing, Fudan University, Shanghai 200433, China}

\begin{abstract}
The new field of spin cavitronics focuses on the interaction between the magnon excitation of a magnetic element and the electromagnetic wave in a microwave cavity.
In strong interaction regime, such interaction usually gives rise to the level anti-crossing for the magnonic and the electromagnetic mode. Recently, the attractive level crossing has been observed, and is explained by a non-Hermitian model Hamiltonian. However, the mechanism of the such attractive coupling is still unclear.
Here we reveal the secret by using a simple model with two harmonic oscillators coupled to a third oscillator with large dissipation. We further identify this dissipative third-party as the invisible cavity mode with large leakage in the cavity-magnon experiments. This understanding enables designing dissipative coupling in all sorts of coupled systems.
\end{abstract}

\maketitle
\end{CJK*}

{\it Introduction.} Spin cavitronics is a newly developing inter-discipline field that combines the spintronics with the cavity quantum electrodynamics, and the purpose is to realize quantum information processing via photon-magnon interaction. The strong interaction between the Kittel mode of magnetic YIG (Yttrium Iron Garnet) sphere \cite{kittel_theory_1948} and the cavity photons has been observed \cite{soykal_strong_2010,huebl_high_2013,zhang_strongly_2014}, even in the quantum regime \cite{tabuchi_hybridizing_2014,lachance-quirion_resolving_2017,tabuchi_quantum_2016}.
The spin cavitronic system provides a platform for interfacing magnon with photon. By placing a magnetic element in a cavity, it is possible to convert optical photon to microwave photon bidirectionally through ferromagnetic magnons \cite{hisatomi_bidirectional_2016}, or transfer spin information between magnet via cavity photons \cite{bai_spin_2015,bai_cavity_2017}.
In the presence of several YIG spheres, the indirect coupling among them can be induced by the cavity photons \cite{lambert_cavity-mediated_2016}, leading to the hybridized magnonic modes \cite{zhang_magnon_2015,xiao_magnon_2019}. These hybrid modes can be interpreted using the molecular orbital theory, so that the design of magnonic molecules with novel properties is expected \cite{zare_rameshti_indirect_2018}. The peculiar dynamics of these modes (bright mode and dark mode) are beneficial for quantum information manipulation and storage \cite{zhang_magnon_2015}.

This mode hybridization between the magnon and photon has no difference from any other coupling systems, where avoided crossing between the energy levels of two eigen-modes is expected, and the size of the anti-crossing gap is proportional to the strength of the coupling.
However, recent experiments have observed the attractive level crossing, in both Fabry-Perot-like cavity \cite{harder_level_2018} and coplanar waveguide-based resonator structure  \cite{bhoi_abnormal_2019,yang_control_2019}. Model Hamiltonians with non-Hermitian dissipative terms have been proposed to interpret the experiments \cite{yao_microscopic_2019}.
Grigoryan \textit{et al.} \cite{grigoryan_synchronized_2018} even constructs an artificial coupling circuit for such non-Hermitian system, which was also demonstrated experimentally \cite{boventer_control_2019}. Such mathematical construction can indeed reproduce the level attraction, but they all lack a physical explanation or mechanism, leaving the level attraction in such systems yet to be understood.

In this Letter, we start with a simple toy model with two harmonic oscillators, which can be coupled via a mutual force proportional to i) their position difference (equivalent to a normal spring) or ii) their velocity difference (no conventional analogy). It can be shown that the type i) coupling is reactive and gives rise to the usual repulsive level crossing, while the type ii) coupling is dissipative and leads to the attractive level crossing. The main contribution of this Letter is to show that in a physical system, the dissipative type ii) coupling can be realized by coupling both oscillators reactively to a third highly dissipative entity. We further identify what is the third-party in the level-attraction experiments, which turned out to be the invisible cavity modes that has extremely high leakage or dissipation.


\begin{figure*}[t]
\centering
\includegraphics[width=0.95\textwidth]{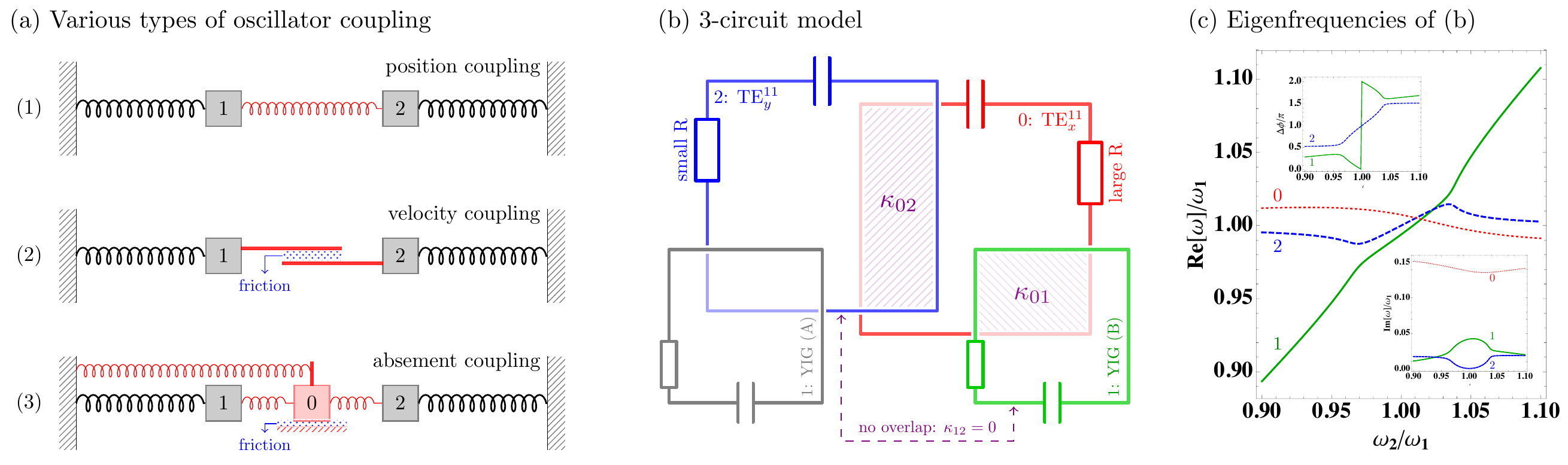}
\caption{The harmonic oscillators with reactive coupling via a direct spring (a-1), dissipative coupling via viscous forces (a-2), and dissipative coupling via a third oscillator $m_0$ in contact with the friction surface (a-3). (b). The 3-circuit model with three RLC circuits: both circuit-2 (blue) and circuit-1 (green) are reactivly coupled to a highly dissipative third-party circuit-3 (red), realizing dissipative coupling between circuit-1 and -2. The circuit-0, -1, -2 here are equivalent to the high dissipation TE$^{11}_x$ mode, the YIG magnon mode, and the cavity TE$^{11}_y$ mode in \Figure{fig:circular}. (c). The eigen frequencies for the 3-circuit model with $\kappa_{12} = 0, \kappa_{0i}/\omega_1^2 = 0.21$, $\gamma_0/\omega_1 = 0.35$, and $\gamma_1/\omega_1 = \gamma_2/2\omega_1 = 0.001$. Inset: the imaginary part of the eigen frequencies (bottom) and the relative phase (top) between circuit-$1$ and -$2$.}
\label{fig:classical_model}
\end{figure*}

{\it The oscillator model.} Let's consider the simplest coupled harmonic oscillator model as shown in \Figure{fig:classical_model}(a). The two oscillators may refer to any physical eigen-modes, and for the present interest of spin cavitronics they can be understood as the cavity photon mode and the Kittel magnon mode in the YIG sphere. Let $\omega_i, \eta_i$ be resonance frequency and damping constant for oscillator-$i$ ($i = 1, 2$). The dynamics of displacements $x_{i=1,2}$ are described as coupled damped oscillators ($i'\equiv 3-i$):
\begin{equation}
  \label{eqn:2ho}
  \ddot{x}_i + \omega_i^2 x_i + \eta_i \dot{x}_i = \kappa_i \hat{T}(x_{i'} - x_i). \\
\end{equation}
The terms on the right hand side represent the generic coupling forces, which are related to the relative displacement $x_2 - x_1$ via an operator $\hat{T}$ and characterized by strength $\kappa_i$. The exact nature of the coupling is represented by the operator $\hat{T}$. For example, for $\hat{T} = \hat{T}_0 = 1$, the coupling can be simply realized by a spring connecting the two oscillators as in \Figure{fig:classical_model}(a-1), \ie the mutual force depends on their relative displacement. If $\hat{T} = \hat{T}_1 = d/dt$ is the time derivative operator, then the mutual force is proportional to their relative velocity, which can be realized via viscous force between the two oscillators as in \Figure{fig:classical_model}(a-2). When $\hat{T} = \hat{T}_{-1} = \int dt$ is the temporal integration operator, then the coupling force is proportional to the relative absement (the time-integral of displacement), which can be realized via a third oscillator with extra dissipation as in \Figure{fig:classical_model}(a-3). In general, we may define $\hat{T}_n \equiv d^n/dt^n$ and $\hat{T}_{-n} \equiv \int d^nt$.
Since for even $n$, $\hat{T}_n$ are even under time reversal, they will lead to reactive coupling. While for odd $n$, $\hat{T}_n$ are odd under time reversal, therefore they represent dissipative coupling.
Solving \Eq{eqn:2ho} by assuming the harmonic ansatz: $x_i(t) = \td{x}_i e^{i\omega t}$, we find that the energy levels (see Supplementary Materials) show the usual repulsive anti-crossing for the reactive couping $\hat{T}_0$, but attractive level crossing for the velocity or absement couplings $\hat{T}_{\pm 1}$. One example for the velocity coupling is proposed in spintronic system where two ferromagnetic layers coupled via spin pumping and spin-transfer torque \cite{heinrich_dynamic_2003}, where the spin current pumped from one magnet acts as spin-transfer torque on the other magnet, and resulting in level attraction or synchrolization.



{\it The 3-circuit model.} To better capture the physics in the cavity-magnonic system, we construct a model system consisting of three mutually overlapping RLC circuits as shown in \Figure{fig:classical_model}(b), equivalent to the previous 3-oscillator model. The coupling in the RLC circuits is through the mutual inductance because of the magnetic field flux threading each other, and the coupling strength $\kappa_{ij} = \kappa_{ji}$ between circuit-$i$ and circuit-$j$ is proportional to the area of their overlapping region. Let the resistance, inductance, and capacitance be denoted by $R_{0,1,2}, L_{0,1,2}, C_{0,1,2}$, respectively, the currents $I_{i=1,2}$ and $I_0$ in the three circuits satisfy:
\begin{subequations}
\label{eqn:3RLC}
\begin{align}
  \ddot{I}_i + \omega_i^2 I_i &= - \gamma_i\dot{I}_i
  + \kappa_{ii'} \ddot{I}_{i'}
  + {\kappa_{0i} \ov\lambda_i} \ddot{I}_0 , \label{eqn:Ii}\\
  \ddot{I}_0 + \omega_0^2 I_0 &= - \gamma_0\dot{I}_0
  + \sum_{j=1,2}\lambda_j \kappa_{0j} \ddot{I}_j, \label{eqn:I0}
\end{align}
\end{subequations}
where $\omega_i = 1/\sqrt{L_iC_i}, \gamma_i = R_i/L_i, \lambda_i = L_i/L_0$. The $\ddot{I}_{i'}$ term in \Eq{eqn:Ii} is the direct coupling between circuit-1 and -2, which is the even (reactive) $\hat{T}_2$ type coupling.

When the circuit-$0$ has small dissipation (small $\gamma_0$), the realized coupling between the circuit-$1$ and -$2$ is the conventional repulsive coupling. However, we shall see below that, if the dissipation of circuit-$0$ is large, the effective coupling between ciruit-$1$ and -$2$ becomes dissipative attractive coupling. To see this point, let's take a limiting case with the left hand side of \Eq{eqn:I0} neglected, then $\ddot{I}_0$ can be replace by $\dddot{I}_{1,2} \sim \hT_3 I_{1,2}$ and \Eq{eqn:Ii} becomes
\begin{equation}
  \label{eqn:Ii2}
  \ddot{I}_i + \omega_i^2 I_i + \gamma_i\dot{I}_i - {\kappa_{0i}^2\ov\gamma_0}\dddot{I}_i
  = \qty(\kappa_{12} \hat{T}_2
  + {\kappa_{01}\kappa_{02}\ov\gamma_0}
  {\lambda_{i'}\ov\lambda_i}\hat{T}_3) I_{i'},
\end{equation}
where the right hand side contains both reactive ($\hat{T}_2$) and dissipative ($\hat{T}_3$) couplings. If $\kappa_{12} = 0$ (no overlap between circuit-1 and -2), the reactive coupling is turned off, and the dissipative coupling dominates. The neglect of the left hand side of \Eq{eqn:I0} can be achieved when the circuit-$0$ oscillates at a frequency close to the resonance frequency: $\omega_0\simeq \omega$ and the dissipation $\gamma_0$ is larger than the detuning: $\gamma_0\omega\gg \abs{\omega^2 - \omega_0^2}$, \footnote{Another condition for negligible left hand side of \Eq{eqn:I0} is to let $L_0\ra 0, C_0\ra\infty$, so that $\gamma_0\dot{I}_0$ dominates \Eq{eqn:I0}.}. Thererfore, to enhance the dissipative coupling (governed by the prefactor of $\hat{T}_3$), one needs to have the third-party dissipation $\gamma_0$  as small as possible, yet large enough such that $\gamma_0\omega \gg \abs{\omega^2 - \omega_0^2}$. \Figure{fig:classical_model}(c) shows such attractive coupling realized in the 3-circuit model in the regime described above.

\begin{figure*}[t]
\centering
\includegraphics[width=0.95\textwidth]{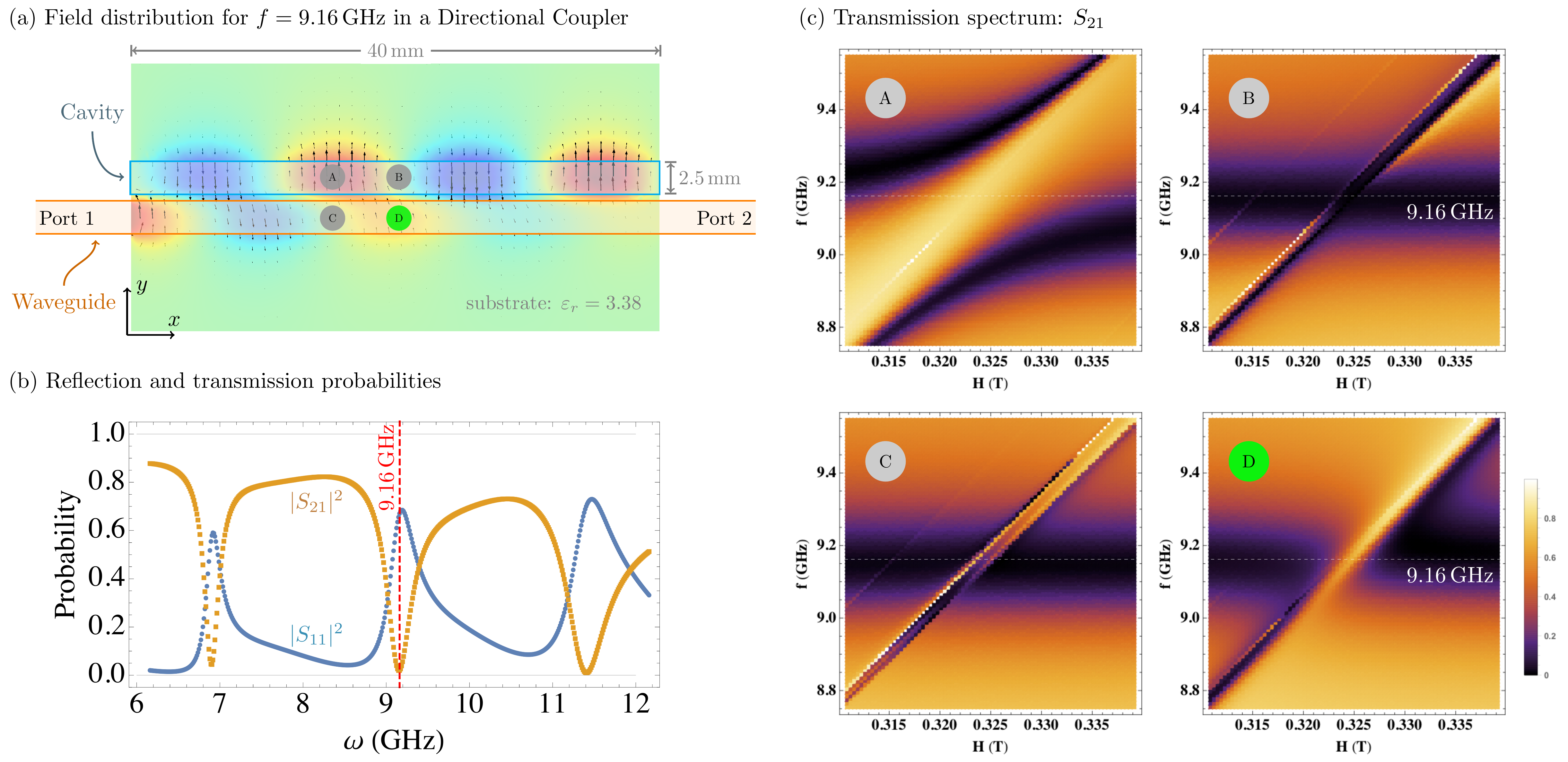}
\caption{(a) The magnetic field distribution for the $f = \SI{9.16}{GHz}$ cavity mode in a directional coupler. The length of the coupler is \SI{40}{mm}, the width of each stripe is \SI{2.5}{mm}, and the relative dielectric constant of the substrate is $\varepsilon_r = 3.38$. (b) The simulated reflection and transmission probabilities $S_{11}$ and $S_{21}$ for the coupler without YIG sphere. (c) The simulated transmission spectrum when the YIG sphere is placed at point A/B/C/D, respectively. All material parameters are the same as those in Ref. \onlinecite{yang_control_2019}.}
\label{fig:dc}
\end{figure*}

{\it The 3-mode Quantum Model.} In a quantum description, using the annihilation operator $\ha_j$ for mode-$j$ ($j=0,1,2$), the system Hamiltonian for three coupled modes can be written as
\begin{align}
  \label{eqn:HH}
  \hH &= \sum_{j=0,1,2} \hbar\omega_j \ha_j^\dagger \ha_j + \sum_{j<k} \kappa_{jk}\qty( \ha_j^\dagger \ha_k + \ha_k^\dagger\ha_j) \\
  &= \mqty(
  \omega_1 & \kappa_{12} & \kappa_{01}\\
  \kappa_{12} & \omega_2 & \kappa_{02} \\
  \kappa_{01} & \kappa_{02} & \omega_0)
  \rightarrow
  \hU^\dagger\hH\hU = \mqty(
  \omega'_1 & \kappa'_{12} & 0\\
  {\kappa'_{12}} & \omega'_2 & 0\\
  0 & 0 & \omega'_0). \nonumber
\end{align}
Since we are interested in the subsystem with mode-$1$ and -$2$, we need to transform away the coupling with the mode-$0$ from the Hamiltonian. This is equivalent to perform a unitary transformation $\hU$ to block diagonalize the 3-mode system into the decoupled 2-mode and 1-mode subsystems as above. When the coupling with the mode-$0$ is weak, by using Schrieffer-Wolff transformation (see Supplementary Materials), we derived the effective coupling between mode-$1$ and -$2$ as
\begin{equation}
  \label{eqn:H2}
  \kappa'_{12} = \kappa_{12} + \half \kappa_{01}\kappa_{02}\sum_{i=1,2} {1\ov \omega_i-\omega_0}.
\end{equation}
and $\omega'_i = \omega_i + \kappa^2_{0i}/(\omega_i-\omega_0)$. When extending to the dissipative regime by allowing complex eigenfrequencies, it is straightforward to see that when the mode-$0$ is extremely dissipative, $\omega_0 = \omega_0^r + i\gamma_0$, the in-direct coupling term becomes imaginary and the Hamiltonian is non-Hermitian. This leads to the level attraction behavior. The condition for the strength of the dissipation $\gamma_0$ is the same as in the classical model in \Eq{eqn:Ii2}, \ie the $\gamma_0$ should be large such that $\omega_i-\omega_0$ is dominated by its imaginary part, but not so large such that the overall strength is still finite.

{\it Level attraction in a directional coupler.}
To illustrate the attractive coupling principle more clearly, we demonstrate the attractive behavior in a directional coupler by simulation (see Supplementary Materials). A directional coupler, as shown in \Figure{fig:dc}(a), consists of two parallel metal stripes on a dielectric substrate. The lower stripe (working as a waveguide) is connected with input and output ports. When the microwave travels through the lower stripe, it partially leaks (couples) to the upper stripe (the cavity), exciting the cavity modes of the upper stripe, which can be detected via the transmission spectrum. The simulated transmission spectrum of the directional coupler, \Figure{fig:dc}(b), shows peaks at the resonance frequencies of the cavity.

Different from the circular/cross cavity used in Ref.\onlinecite{harder_level_2018} and \onlinecite{yang_control_2019}, the cavity and the dissipative mode in a directional coupler are spatially separated, which enables us to see the coupling more clearly. Let's focus on the cavity mode at $f=\SI{9.16}{GHz}$. \Figure{fig:dc}(a) shows the spatial distribution of the in-plane magnetic field of this mode over a plane slightly beneath the metal stripes. The cavity mode in the upper stripe has vanishing  magnetic field at the edges and at the center of the stripe. While the dissipative mode in the lower stripe has maximum magnetic field at the edges and at the center, which has a quarter wavelength offset from the cavity mode. This offset between the cavity and the dissipative mode means that the locations with the maximum magnetic field for the cavity mode correspond to the spots of zero magnetic field for the dissipative mode, and vice versa.

We now study the transmission spectrums when the YIG sphere (with diameter \SI{1}{mm}) is placed 0.75 mm beneath the metal stripe and at four different locations A/B/C/D. Point A is the magnetic field antinode (of maximum field) for the cavity mode, therefore the YIG sphere is coupled to the cavity directly, resulting in the conventional repulsive coupling (see panel A in \Figure{fig:dc}(c)). While points B and C are on the magnetic field node (of zero or negligible field) for cavity and the dissipative modes, thus no coupling is observed (panel B/C in \Figure{fig:dc}(c)). Point D, however, is the magnetic field node for the cavity mode, but antinode for the dissipative mode, therefore the YIG sphere is coupled only to the dissipative mode but not to the cavity mode. Considering that the cavity mode is also coupled to the dissipative mode along the path, the scenario of YIG and cavity modes coupled via a third dissipative mode is realized, leading to attractive level crossing as shown in panel D of \Figure{fig:dc}(c).


\begin{figure}[t]
\centering
\includegraphics[width=0.98\columnwidth]{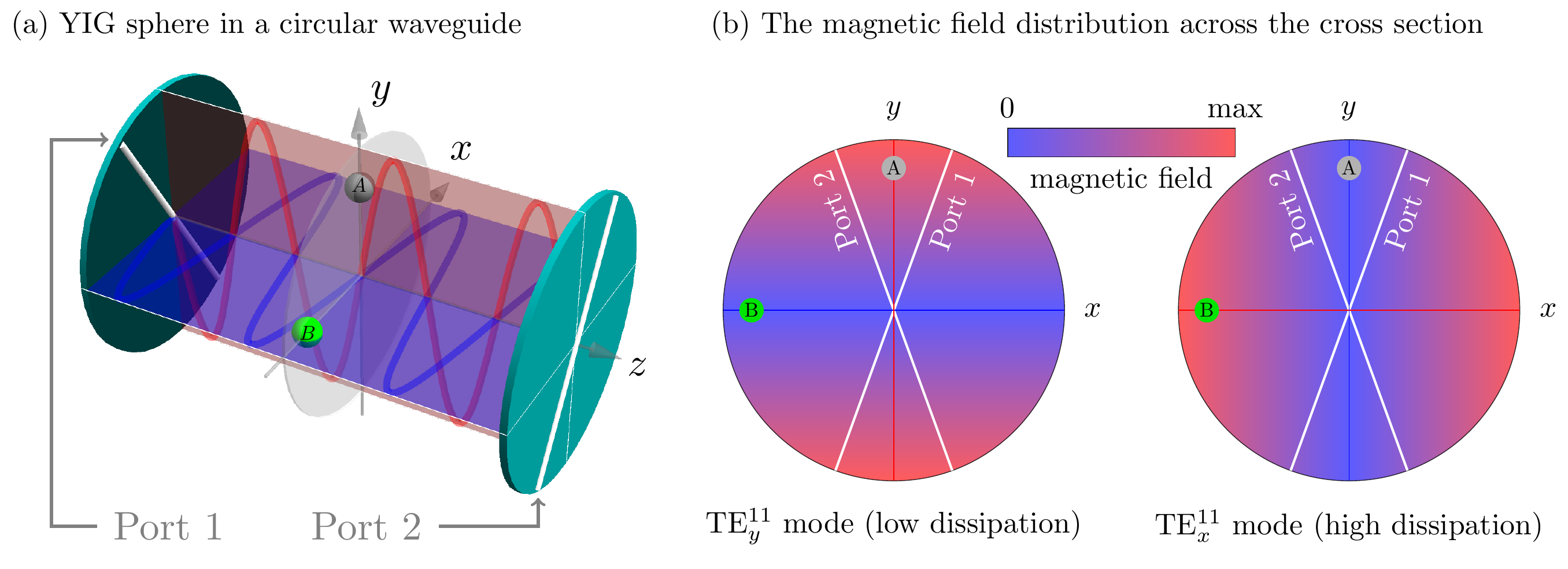}
\caption{The magnetic field vanishes along $\hbx$ for the cavity TE$^{11}_y$ mode (left) and vanishes along $\hby$ for the TE$^{11}_x$ mode (right). When the YIG sphere is placed at point A (B), it couples strongly with TE$^{11}_x$ (TE$^{11}_y$) mode.}
\label{fig:circular}
\end{figure}

{\it Interpretation of the level attraction in the circular waveguide.} In the circular cavity-YIG system (schematically illustrated in \Figure{fig:circular}(a)) studied in Ref.\onlinecite{harder_level_2018}, the repulsive and attractive level crossing are obeserved when the YIG sphere is placed at point A and B, which correspond to the magnetic field antinode and node for the cavity mode respectively. We shall be clear below that the cavity and the YIG mode realize attractive level crossing via their coupling to a common dissipative third-party, just as in the cases studied above. The issue is what the third-party is in the circular cavity? To answer that, we note that the circular cavity of interest has two ports, forming a small angle (see \Figure{fig:circular}(a)). These ports play two roles: i) they are used for feeding and draining the TE electromagnetic wave into and out of the cavity, ii) they reflect TE waves polarized perpendicular to the port orientations.
If there were no port, the cavity should have EM modes of all polarizations. However, because of the opening ports, the cavity mode with polarization along $\hby$ (call it TE$^{11}_y$) has the longest lifetime. And the EM mode polarized along $\hbx$ (TE$^{11}_x$) has the largest leakage through the two ports, thus the shortest lifetime. The long-lived TE$^{11}_y$ mode is the visible cavity mode we usually measure, and its magnetic field (anti-)node is along $\hbx$ ($\hby$) (see \Figure{fig:circular}(b)). The TE$^{11}_x$ mode is invisible due to its extremely short lifetime, and its magnetic field (anti-)node is opposite to that of TE$^{11}_y$ mode (see \Figure{fig:circular}(b)). Therefore, when the YIG sphere is placed at point B, it is not coupled to the cavity mode TE$^{11}_y$, but coupled to the dissipative TE$^{11}_x$ mode. In addition, the TE$^{11}_x$ and TE$^{11}_y$ modes are coupled through reflections by the ports. Consequently, the TE$^{11}_x$ works as the dissipative third-party that couples the cavity TE$^{11}_y$ mode and the YIG magnon mode together, leading to their attractive level crossing as observed in the experiment.


The high-dissipation TE$^{11}_x$ mode satisfies the requirement for working as the third-party: the TE$^{11}_x$ mode has the same eigen frequency as the cavity TE$^{11}_y$ mode, so $\omega_0 \simeq \omega_1$ is very close to the resonant frequency $\omega$. This minimizes $\abs{\omega^2 - \omega_0^2}$. Because the TE$^{11}_x$ mode has high leakage through the ports (thus large $\gamma_0$), $\gamma_0\omega \gg \abs{\omega^2 - \omega_0^2}$ is naturally satisfied.

{\it Discussion \& Conclusion}. The level attraction via a dissipative mode is a general physical principle, which can be applied to a wide range of coupled physical systems. For example, either the oscillator or the dissipative third-party can be superconducting qubit \cite{tabuchi_coherent_2015,lachance-quirion_resolving_2017}, dielectric nanostructures \cite{preston_vibron_2011}, antiferromagnets \cite{johansen_nonlocal_2018,yuan_magnon-photon_2017}, high-order spin wave modes \cite{cao_exchange_2015} or other excitations such as phonons \cite{zhang_cavity_2016}. It has been reported recently that magnetic textures can also be coupled with the cavity photons  \cite{proskurin_cavity_2018,abdurakhimov_magnon-photon_2019}. Based on the understanding of dissipative coupling, the nonlinear effect \cite{wang_magnon_2016,wang_bistability_2018} and topological properties of exceptional point \cite{harder_topological_2017,zhang_observation_2017,zhang_higher-order_2019,cao_exceptional_2019} can be generalized and new physics is expected.

In conclusion, we found that the mechanism for the dissipative coupling in many physical systems can be captured by an effective 3-oscillator model, where two oscillators of interests are coupled to a common third oscillator with strong dissipation. We verify this model in both classical and quantum setup. Based on this model, we are able to explain the exact physical mechanisms behind the level attraction experiments carried out in the cavity-magnon systems, where a hidden cavity mode with large dissipation is responsible for mediating the dissipative coupling.

{\it Acknowledgements.}
This work was supported by the National Natural Science Foundation of China (No. 11722430, 11847202, 61704071).
W.Y. is supported by the China Postdoctoral Science Foundation under Grant No. 2018M641906.



\bibliography{ref}

\end{document}